\documentclass{article}
\usepackage{graphicx}
\usepackage{caption}
\usepackage{subcaption}
\usepackage[dvipsnames]{xcolor}
\usepackage{amsmath}
\usepackage{amsfonts}
\usepackage{braket}
\usepackage{amsthm}
\usepackage{amssymb}
\usepackage{hyperref}
\usepackage{tikz}
\usepackage{tikzscale}
\usepackage{algorithm}
\usepackage{algpseudocode}
\usepackage[inline]{enumitem}
\usepackage[a4paper, total={6in, 8in}]{geometry}
\usetikzlibrary{decorations.pathreplacing}
\usetikzlibrary{calc}
\usetikzlibrary{quantikz2}
\usetikzlibrary[graphs, graphs.standard]
\usetikzlibrary{positioning,chains,fit,shapes,calc,decorations.markings}
\usetikzlibrary{decorations.pathreplacing,calligraphy}

\newcommand{\black}[1]{{\color{black} #1}}
\newcommand{\white}[1]{{\color{White} #1}}

\newcommand{\Tau}{\mathcal{T}}
\newcommand{\sz}{SzQW}
\newcommand{\st}{StQW}
\newcommand{\abs}[1]{\left|#1\right|}

\renewcommand{\P}{\mathcal{P}}
\newcommand{\R}{\mathcal{R}}
\newcommand{\W}{\mathcal{W}}

\newcommand{\I}{\mathcal{I}}
\newcommand{\A}{\mathcal{A}}
\newcommand{\B}{\mathcal{B}}
\newcommand{\G}{\mathcal{G}}
\newcommand{\D}{\mathcal{D}}
\newcommand{\F}{\mathcal{F}}
\newcommand{\pr}[1]{\left( #1 \right)}
\renewcommand{\proj}{\Pi}
\newcommand{\diag}[1]{\textnormal{diag}\pr{#1}}
\newcommand{\im}{\textnormal{i}}
\newcommand{\hilb}{\mathcal{H}}
\renewcommand{\set}[1]{\left\{ #1 \right\}}
\newcommand{\cdb}{CDB}
\newcommand{\qdb}{QDB}
\newcommand{\eps}{\varepsilon}
\newcommand{\ceil}[1]{\left\lceil #1 \right\rceil}
\renewcommand{\max}{{\textnormal{max}}}
\newcommand{\norm}[1]{\abs{\abs{#1}}}
\newcommand{\ambainis}{AGJK}
\renewcommand{\emptyset}{\varnothing}

\newtheorem{theorem}{Theorem}[]
\newtheorem{proposition}[theorem]{Proposition}
\newtheorem{lemma}[theorem]{Lemma}
\newtheorem{corollary}[theorem]{Corollary}

\title{Quantum Search on Bipartite Multigraphs}
\makeatletter
\renewcommand\@date{{%
  \vspace{-\baselineskip}%
  \large\centering
  \begin{tabular}{@{}c@{}}
    Gustavo Alves Bezerra\textsuperscript{1} \\
    \normalsize gbezerra@posgrad.lncc.br
  \end{tabular}%
  \quad
  \begin{tabular}{@{}c@{}}
    Andris Ambainis\textsuperscript{2} \\
    \normalsize andris.ambainis@lu.lv
  \end{tabular}
  \quad
  \begin{tabular}{@{}c@{}}
    Renato Portugal\textsuperscript{1} \\
    \normalsize portugal@lncc.br
  \end{tabular}

  \bigskip

  \textsuperscript{1}National Laboratory of Scientific Computing,
    Brazil\par
  \textsuperscript{2}Center for Quantum Computer Science,
    University of Latvia, Latvia

  \bigskip

  \today
}}
\makeatother

\begin{document}

\maketitle

\begin{abstract}
Quantum walks provide a powerful framework for achieving algorithmic speedup in quantum computing. This paper presents a quantum search algorithm for 2-tessellable graphs, a generalization of bipartite graphs, achieving a quadratic speedup over classical Markov chain-based search methods. Our approach employs an adapted version of the Szegedy quantum walk model (adapted SzQW), which takes place on bipartite graphs, and an adapted version of Staggered Quantum Walks (Adapted StQW), which takes place on 2-tessellable graphs, with the goal of efficiently finding a marked vertex by querying an oracle. The Ambainis, Gilyén, Jeffery, and Kokainis' algorithm (\ambainis), which provides a quadratic speedup on balanced bipartite graphs, is used as a subroutine in our algorithm. Our approach generalizes existing quantum walk techniques and offers a quadratic speedup in the number of queries needed, demonstrating the utility of our adapted quantum walk models in a broader class of graphs.
\end{abstract}

\section{Introduction}

Quantum walks are the quantum counterparts of classical random walks~\cite{aharonov1993quantum} and constitute a universal model of computation~\cite{childs2009universal, lovett2010universal}. They have been applied to problems such as element distinctness~\cite{ambainis2007quantum} and spatial search~\cite{benioff2002space, shenvi2003quantum}. In the spatial search problem on a graph with $N$ vertices, an oracle marks the desired vertex or vertices, and the goal is to locate one of the marked vertices with the fewest possible queries to the oracle. Spatial search has been studied using various quantum walk models, including coined~\cite{aharonov2001quantum}, Szegedy's~\cite{szegedy2004quantum}, continuous-time~\cite{farhi1998analog}, and staggered~\cite{portugal2016stqw}.

In the coined model, if we consider that the walker is located at a vertex, the coin determines the distribution of amplitudes to neighboring vertices using the graph structure~\cite{aharonov2001quantum}. Coined quantum walks were initially applied to spatial search, achieving a query complexity of $O(\sqrt N)$ for the hypercube~\cite{shenvi2003quantum} and $\tilde O(\sqrt N)$ for the two-dimensional grid~\cite{ambainis2005coins}.\footnote{$\tilde O(\cdot)$ notation hides polylogarithmic factors.}~ Some polylogarithmic factors were eliminated by adding ancillas~\cite{tulsi2008}, using classical post-processing~\cite{ambainisnahimovs20213}, and increasing laziness~\cite{wong2018faster}. The problem of quantum search with multiple marked vertices has also been investigated~\cite{ambainis2008quantum, abhijith2018spatial, bezerra2021quantum}. However, in the coined model, certain cases involving multiple marked vertices result in search failure~\cite{ambainis2008quantum, prusis2016stationary}.

Szegedy~\cite{szegedy2004quantum} introduced a quantum walk model based on Markov chains. In this model, the Markov chain is transformed into a balanced bipartite graph (parts of equal size), and the quantum walk takes place in the Hilbert space spanned by the edges of this bipartite graph. Szegedy applied this framework to the detection problem. Krovi et al.~\cite{krovi2016quantum} proposed a quantum algorithm based on interpolated quantum walks to locate marked vertices on a Markov chain, achieving a query complexity of $O(\sqrt{HT^+})$\footnote{$HT^+$ denotes the extended hitting time.}. Ambainis et al.~\cite{ambainis2020quadratic} later demonstrated that $HT^+$ can be significantly larger than the hitting time $HT$ when multiple marked vertices are present. In the same work, they proposed a quantum algorithm (\ambainis's algorithm) for searching a marked vertex on a Markov chain with a query complexity of $\tilde O(\sqrt{HT})$ even for multiple marked vertices. Our goal in this paper is to extend \ambainis's algorithm to a larger class of graphs.

In the continuous-time quantum walk model, the evolution operator is derived from the adjacency or Laplacian matrix of the graph, which is implemented as a Hamiltonian~\cite{farhi1998quantum,childs2004spatial}. For spatial search, Farhi and Gutmann~\cite{farhi1998analog} achieved a query complexity of $O(\sqrt{N})$ on the complete graph, while Childs and Goldstone~\cite{childs2004spatial} demonstrated a query complexity of $\tilde O(\sqrt N)$ for the $d$-dimensional grid with $d \geq 4$. Apers et al.~\cite{apers2022quadratic} applied a technique similar to that in~\cite{ambainis2020quadratic} to solve quantum search with $\tilde O(\sqrt{HT})$ queries for graphs corresponding to reversible Markov chains. Using an asymptotic approach, Lugão et al.~\cite{lugao2024multimarked} achieved a query complexity of $O(\sqrt N)$ for Johnson graphs, while Lugão and Portugal~\cite{lugao2024quantum} demonstrated the same complexity for certain $t$-designs.

The staggered quantum walk model, introduced by Portugal et al.~\cite{portugal2016stqw,portugal2016stqwongraphs}, is based on graph tessellations and extends Szegedy's quantum walks~\cite{portugal2016equivalence}. Portugal et al.~\cite{portugal2017search-hamiltonians} later generalized the evolution operator of staggered quantum walks by incorporating Hamiltonians. Initially, when staggered quantum walks without Hamiltonians were applied to the search problem on the two-dimensional grid, the results were no better than those of classical algorithms. However, by incorporating Hamiltonians into the evolution operator, Portugal and Fernandes~\cite{portugal2017search-hamiltonians} successfully performed quantum search on a grid and other general structures. More recently, Higuchi et al.~\cite{higuchi2019eigenbasis} derived the eigenbasis of the evolution operator for staggered quantum walks with Hamiltonians on 2-tessellable graphs, interpreting the evolution operator as a quantum Markov chain.

The set of 2-tessellable graphs corresponds to the line graph of the set $\mathcal{S}$ of bipartite multigraphs, meaning that the set of 2-tessellable graphs is given by $L(\mathcal{S})$, where $L$ is the line-graph operator~\cite{abreu2020}.
Given a graph $G$, its line graph $L(G)$ is defined such that each vertex of $L(G)$ corresponds to an edge of $G$, and two vertices in $L(G)$ are adjacent if and only if their corresponding edges in $G$ share a common endpoint.
Conversely, if $\mathcal{S}'$ represents the set of 2-tessellable graphs, then the set $K(\mathcal{S}')$ corresponds to the set of bipartite multigraphs, where $K$ is the clique-graph operator. For any graph $G'$, its clique graph $K(G')$ is defined as the graph whose vertices correspond to the maximal cliques of $G'$, with two vertices adjacent if their corresponding cliques share at least one vertex. Moreover, if the cliques share more than one vertex, then $K(G')$ contains multiedges.

A key limitation of \ambainis's algorithm is that it requires a balanced bipartite graph. Our goal is to extend quantum search to a broader class of graphs: bipartite multigraphs. To achieve this, we first generalize Szegedy's quantum walk model to arbitrary bipartite graphs (balanced or unbalanced). We then adapt this model to perform quantum search, allowing for arbitrary sets of marked vertices. Lastly, we adapt the staggered quantum walk model to perform quantum search on 2-tessellable graphs allowing for arbitrary sets of marked cliques.
By formulating the search in terms of the clique graph of 2-tessellable graphs, our approach effectively performs quantum search on bipartite multigraphs, allowing for arbitrary sets of marked vertices.

In the standard Szegedy's model, a Markov chain is first defined on a simple graph, from which a balanced bipartite graph is derived. In contrast, our approach starts with a quantum walk on an arbitrary bipartite multigraph with marked vertices, from which an underlying reversible Markov chain is obtained. This enables us to apply the \ambainis's algorithm~\cite{ambainis2020quadratic} as a subroutine. However, since the underlying Markov chain does not contain sufficient information to uniquely reconstruct the original bipartite multigraph, the algorithm must take as input the bipartite multigraph itself, along with an oracle that identifies the marked vertices. If the underlying Markov chain has a hitting time~$HT$, the adapted \sz\ finds a marked vertex on any bipartite graph with $\tilde O(\sqrt{HT})$ queries to the oracle,
and the adapted \st\ finds a marked clique on any 2-tessellable graph with $\tilde O(\sqrt{HT})$ queries.
Moreover, by interpreting the search in terms of the clique graph of a 2-tessellable graph, our approach can be understood as finding a marked vertex on any bipartite multigraph with $\tilde O(\sqrt{HT})$ queries to the oracle.

This paper is organized as follows. Section~\ref{sec:framework} presents key definitions related to Markov chains and reviews relevant prior work. Section~\ref{sec:lsz} introduces adapted Szegedy's quantum walks and demonstrates their application to quantum search. Section~\ref{sec:lst} introduces the adapted staggered quantum walk and shows its effectiveness for quantum search. Finally, in Section~\ref{sec:conclusion}, we summarize our results and provide a brief discussion.

\section{Preliminaries}
\label{sec:framework}
\noindent
This section reviews key concepts related to Markov chains, followed by a discussion on Szegedy's quantum walk model and its connection to these chains. The concept of interpolated quantum walks is then introduced, providing the foundation for \ambainis's algorithm, which
uses these interpolations for efficient quantum search.
Lastly, the staggered quantum walk model is reviewed, emphasizing its relationship with Szegedy's approach and highlighting its potential advantages.

\subsection{Discrete-time Markov chain}
\label{sec:markov-chain}
\noindent
A discrete-time Markov chain consists of a random walk on a finite simple graph $G(V,E)$, where $V$ is the vertex set and $E$ is the edge set. The transition matrix of the Markov chain is a $|V|$-dimensional matrix $P$, where $P_{vu} > 0$ if $uv \in E$,
and $P_{vu} = 0$ otherwise. Moreover,
\begin{align}
\sum_{v \in V} P_{vu} = 1,\ \forall u \in V.
\end{align}
Here, $P_{vu}$ represents the probability of transitioning from $u$ to $v$.

A Markov chain $P$ is {ergodic} if it has a unique {stationary distribution} $\vec\pi$ that satisfies $P\vec\pi = \vec\pi$ and, for any initial probability distribution $\vec p$, the sequence $P^t\vec p$ converges to $\vec\pi$ as $t$ becomes sufficiently large. In this case, $P$ has exactly one $+1$-eigenvector and $P^t_{uv}>0$ for all $u$ and $v$ when $t$ is sufficiently large.

The fundamental theorem of Markov chains states that $P$ is ergodic if it is irreducible and aperiodic. This means that for any pair of vertices $u$ and $v$, the walker can go from $u$ to $v$ with nonzero probability, and the greatest common divisor (gcd) of the lengths of all closed walks from $u$ to $u$ is equal to 1. Consequently, a Markov chain defined by $P$ is ergodic if and only if the graph $G(V,E)$ is connected and not bipartite.

The {time-reversed Markov chain} of an ergodic Markov chain $P$ is defined as
\begin{align}
P^{-1} \equiv \diag{\vec\pi}P^T\diag{\vec\pi}^{-1},
\end{align}
where $\diag{\vec\pi}$ is the matrix with the entries of $\vec\pi$ on the main diagonal.

For an ergodic Markov chain $P$, the {discriminant} matrix is defined as
$D_{uv} \equiv \sqrt{P_{vu}P^{-1}_{uv}}$, or alternatively,
\begin{align}
D = \diag{\vec\pi}^{1/2} P\ \diag{\vec\pi}^{-1/2}.
\label{eq:discriminant}
\end{align}

A {reversible Markov chain} is defined as an ergodic Markov chain that satisfies the \emph{Classical Detailed Balance condition} (\cdb),
\begin{align}
P_{vu}\vec\pi_u = P_{uv}\vec\pi_v,
\label{eq:cdb}
\end{align}
where $\vec\pi_u$ are the entries of $\vec\pi$.

In this work, we employ systematically bipartite graphs $G(V,E)$, where $V=V_1\cup V_2$ and $N_1\equiv|V_1|$, $N_2\equiv|V_2|$. The {transition matrix} $P_1 : V_1 \to V_2$ is a column-stochastic matrix, where $(P_1)_{vu}$ corresponds to the probability of transitioning from $u$ to $v$,
and
\begin{align}
    \sum_{v \in V_2} (P_1)_{vu} = 1\ \forall u \in V_1.
\end{align}
We are using~\cite{Bondy2011} as a reference for basic concepts and notation in graph theory.


\subsection{Szegedy's quantum walk}
\label{sec:sz}
\noindent
Szegedy proposed to implement Markov chains in the quantum context using bipartite quantum walks.
Szegedy's Quantum Walk (\sz) is defined using two transition matrices
$P_1 : V_1 \to V_2$ and $P_2 : V_2 \to V_1$ where
$V_1$ and $V_2$ are sets of vertices.
The Hilbert space is $\hilb^{N_1} \otimes \hilb^{N_2}$,
therefore the computational basis state $\ket{uv}$ always imply that
$u \in V_1$ and $v \in V_2$.
We define auxiliary states
\begin{align}
    \ket{\alpha_u} \equiv \sum_{uv \in E} a_{uv} \ket{uv},
\end{align}
and
\begin{align}
    \ket{\beta_v} \equiv\sum_{uv \in E} b_{uv} \ket{uv},
\end{align}
where
$a_{uv} \equiv \sqrt{(P_1)_{vu}}$ is the square root of the probability of going from vertex $u$ to $v$,
and $b_{uv} = \sqrt{(P_2)_{uv}}$ is the square root of the probability of going from vertex $v$ to $u$\footnote{We
    use $a_{uv}$ and $b_{uv}$ to highlight the relationship of Szegedy's and
    staggered quantum walks.}.
We highlight that $\set{\ket{\alpha_u}}$ and $\set{\ket{\beta_v}}$ form two orthonormal bases,
and $D_{uv} = \braket{\alpha_u | \beta_v}$.
We define projectors using the auxiliary states,
$\proj_\alpha \equiv \sum_u \ket{\alpha_u}\bra{\alpha_u}$, and
$\proj_\beta \equiv \sum_v \ket{\beta_v}\bra{\beta_v}$.
The \sz\ evolution operator is given by
\begin{align}
    U \equiv \pr{2 \proj_\beta - I}
        \pr{2 \proj_\alpha - I}.
    \label{eq:sz-evol-op}
\end{align}
This definition using two different transition matrices $P_1$ and $P_2$,
and two different vertices sets will be useful later.

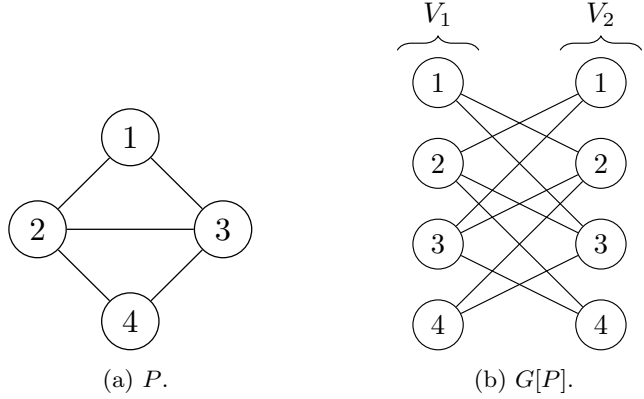
\begin{figure}[htbp]
    \vspace*{13pt}
    \centering
    \begin{subfigure}[t]{0.22\textwidth}
        \centering
        \resizebox{\textwidth}{!}{\begin{tikzpicture}[every node/.style={draw,circle}]
    \node (4) at (0, 0) {4};
    \node (3) at (1, 1) {3};
    \node (2) at (-1, 1) {2};
    \node (1) at (0, 2) {1};

    \draw (1) -- (2); \draw (1) -- (3);
    \draw (2) -- (3); \draw (2) -- (4);
    \draw (3) -- (4);
\end{tikzpicture}}
        \caption{$P$.}
        \label{fig:markov-chain}
    \end{subfigure}
    \hspace{0.1\textwidth}
    \begin{subfigure}[t]{0.22\textwidth}
        \centering
        \resizebox{\textwidth}{!}{\begin{tikzpicture}
    \begin{scope}[every node/.style={draw,circle}]
        \node (4x) at (0, 0) {4};
        \node (3x) at (0, 1) {3};
        \node (2x) at (0, 2) {2};
        \node (1x) at (0, 3) {1};

        \node (4y) at (2, 0) {4};
        \node (3y) at (2, 1) {3};
        \node (2y) at (2, 2) {2};
        \node (1y) at (2, 3) {1};
    \end{scope}
    
    \draw (1x) -- (3y); \draw (1y) -- (3x);
    \draw (1x) -- (2y);
    \draw (1y) -- (2x);
    \draw (2x) -- (3y); \draw (2y) -- (3x);
    \draw (2y) -- (4x); \draw (4x) -- (3y);
    \draw (4y) -- (2x); \draw (4y) -- (3x);

    \draw [decorate, decoration={brace, amplitude=5pt}]
        (-0.5, 3.4) -- (0.5, 3.4)
        node[midway, yshift=1.25em]{$V_1$};
        
    \draw [decorate, decoration={brace, amplitude=5pt}]
        (1.5, 3.4) -- (2.5, 3.4)
        node[midway, yshift=1.25em]{$V_2$};
\end{tikzpicture}}
        \caption{$G[P]$.}
        \label{fig:bip-graph}
    \end{subfigure}
    \vspace*{13pt}
    \caption{Markov chain and its associated bipartite graph.}
\end{figure}

In \sz s, it is assumed that $V = V_1 = V_2$ and $P = P_1 = P_2$.
The Markov chain $P$ induces a balanced bipartite graph $G[P]$ through
a duplication process.
The \sz\ takes place on $G[P]$.
For example, Fig.~\ref{fig:markov-chain} illustrates a Markov chain $P$ where
$P_{uv} \neq 0 \iff P_{vu} \neq 0$ and the weights are omitted.
To obtain $G[P]$,
we add an edge from $u \in V_1$ to $v \in V_2$ and
an edge from $u \in V_2$ to $v \in V_1$
if and only if $P_{uv} \neq 0$
(Fig.~\ref{fig:bip-graph}).

There is an alternative way of implementing the
evolution operator which is going to be very useful to our purposes.
First, we are augment the second register with
a new reference state $\ket 0$.
Second, we define a few auxiliary unitary operators.
\begin{align}
    A \ket{u0} &\equiv \ket{\alpha_u}
\end{align}
is the operator that implements the desired transitions of the Markov chain;
\begin{align}
    S \ket{uv} \equiv \begin{cases}
        \ket{vu} \text{if } v \neq 0, \\
        \ket{u0} \text{if } v = 0
    \end{cases}
\end{align}
is a swap operator, which allows us to obtain
\begin{align}
    B  \ket{v0} = \ket{\beta_v},
\end{align}
where $B\equiv SA$ and
\begin{align}
    R &\equiv 2\proj - I
    \label{eq:ref-operator}
\end{align}
is the reflection operator where
\begin{align}
    \proj \equiv \sum_u \ket{u0} \bra{u0}
\end{align}
is the projector onto the space spanned by $\{\ket{u0}\}$.
Using the auxiliary operators we define,
\begin{align}
    W \equiv B^\dagger A R.
\end{align}
Then, we use the equation above, $B=SA$, and $AA^\dagger=I$ to rewrite the evolution operator as
\begin{align}
    U &= A W^2 A^\dagger \label{eq:original-sz-U}\\
    &= B R B^\dagger A R A^\dagger \\
    &= \pr{2 \proj_\beta - I}\pr{2 \proj_\alpha - I}.
\end{align}

Thus, the evolution of \sz\ and most of the analysis can be described in terms of $W$.
If we restrict $W$ to the subspace spanned by $\set{\ket{u0}}$,
we notice that it implements the discriminant matrix,
\begin{align}
    D_{uv} = \braket{\alpha_u | \beta_v}
    = \bra{u0} W \ket{v0}.
\end{align}

\subsection{Interpolated Szegedy quantum walk}
\label{sec:interpolated-sz}
\noindent
Let $M$ be the set of marked vertices and
$\bar M$ be the set of unmarked vertices.
After a permutation, the Markov chain $P$ can be rewritten as
\begin{align}
    P = \begin{bmatrix}
        P_{\bar M \bar M} & P_{\bar M M} \\
        P_{M \bar M} & P_{M M}
    \end{bmatrix}
\end{align}
where $P_{\bar M M}$ denotes the block matrix corresponding to
transitions from marked to unmarked vertices --
analogous for the remaining blocks.
To perform quantum search on \sz s,
we mark the desired vertices of the Markov chain by turning them into sinks.
In other words,
we remove all transitions leaving the marked vertices and
add self-loops to them if needed.
This gives the Markov chain
\begin{align}
    P' \equiv \begin{bmatrix}
        P_{\bar M \bar M} & 0 \\
        P_{M \bar M} & I
    \end{bmatrix}.
\end{align}

\begin{figure}[!htb]
    \vspace*{13pt}
    \centering
    \begin{subfigure}[t]{0.22\textwidth}
        \centering
        \resizebox{\textwidth}{!}{\begin{tikzpicture}[every node/.style={draw,circle}]
    \node [ultra thick] (4) at (0, 0) {\textbf{4}};
    \node (3) at (1, 1) {3};
    \node (2) at (-1, 1) {2};
    \node (1) at (0, 2) {1};

    \draw (1) -- (2); \draw (1) -- (3);
    \draw (2) -- (3); \draw [-latex] (2) -- (4);
    \draw [-latex] (3) -- (4);
    \path (4) edge [loop above, >=latex] (4);
\end{tikzpicture}}
        \caption{$P'$.}
        \label{fig:marked-markov-chain}
    \end{subfigure}
    \hspace{0.1\textwidth}
    \begin{subfigure}[t]{0.22\textwidth}
        \centering
        \resizebox{\textwidth}{!}{\begin{tikzpicture}
    \begin{scope}[every node/.style={draw,circle}]
        \node [ultra thick] (4x) at (0, 0) {\textbf{4}};
        \node (3x) at (0, 1) {3};
        \node (2x) at (0, 2) {2};
        \node (1x) at (0, 3) {1};

        \node [ultra thick] (4y) at (2, 0) {\textbf{4}};
        \node (3y) at (2, 1) {3};
        \node (2y) at (2, 2) {2};
        \node (1y) at (2, 3) {1};
    \end{scope}
    
    \begin{scope}[every path/.style={-latex}]
        \draw (2y) -- (4x); \draw (3y) -- (4x);
        \draw (2x) -- (4y); \draw (3x) -- (4y);
    \end{scope}

    \draw (1x) -- (2y);
    \draw (1y) -- (2x);
    \draw (1x) -- (3y); \draw (1y) -- (3x);
    \draw (2x) -- (3y); \draw (2y) -- (3x);
    \draw (4y) -- (4x);

    \draw [decorate, decoration={brace, amplitude=5pt}]
        (-0.5, 3.4) -- (0.5, 3.4)
        node[midway, yshift=1.25em]{$V_1$};
        
    \draw [decorate, decoration={brace, amplitude=5pt}]
        (1.5, 3.4) -- (2.5, 3.4)
        node[midway, yshift=1.25em]{$V_2$};
\end{tikzpicture}}
        \caption{$G[P']$.}
        \label{fig:marked-bip-graph}
    \end{subfigure}
    \vspace*{13pt}
    \caption{Markov chain with a sink and its associated bipartite digraph.}
\end{figure}
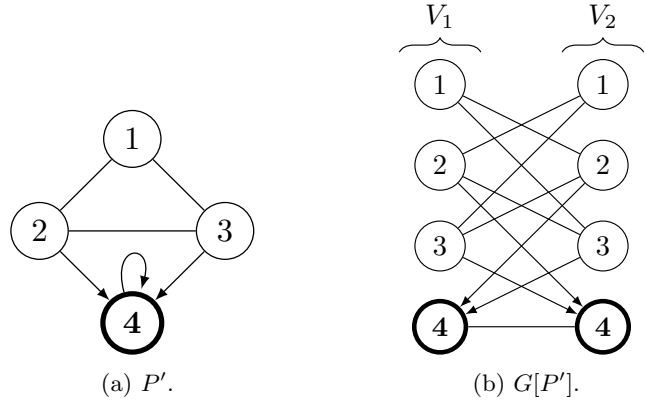

For example,
if we mark vertex $4$ of the Markov chain in Fig.~\ref{fig:markov-chain},
we obtain the Markov chain $P'$ in Fig.~\ref{fig:marked-markov-chain}.
The induced graph $G[P']$ is the directed version of $G[P]$ where
the arcs leaving $4 \in V_1$ and $4 \in V_2$ were removed,
and we add an edge from $4 \in V_1$ to $4 \in V_2$
(Fig.~\ref{fig:marked-markov-chain}).
Note that $\ket{\alpha_4}$ obtained from the original Markov chain $P$ is a superposition of
$\ket{42}$ and $\ket{43}$; while $\ket{\beta_4}$ is a superposition of $\ket{24}$ and $\ket{34}$.
On the other hand, from the marked Markov chain $P'$,
we obtain $\ket{\alpha'_4} = \ket{\beta'_4} = \ket{44}$.

Using $P$ and $P'$ we define the \emph{interpolated Markov chain}
\begin{align}
    P(r) \equiv (1 - r)P + rP'
\end{align}
for any $0 \leq r < 1$.
If $P$ is reversible, then $P(r)$ is reversible with
unique stationary distribution\footnote{Note
    that if we had allowed $r = 1$,
    $P(r)$ would not always be reversible because
    $P'$ has multiple stationary distributions.
    For example, any linear combination of the marked vertices.
}
\begin{align}
    \vec\pi(r) \equiv \frac{(1 - r)\vec\pi + r\vec\pi_M}{(1 - r) + r p_M}
\end{align}
where
\begin{align}
    \vec\pi_{M} \equiv
    \frac{1}{p_M} \sum_{u \in M} \vec\pi_{u} \ket{u}
\end{align}
is the normalized projection of $\vec\pi$ onto the subspace spanned by the marked vertices, and
\begin{align}
    p_M \equiv \sum_{u \in M} \vec\pi_u
\end{align}
is the sum of the probability of all marked vertices in the stationary distribution $\vec\pi$
(analogous for $\vec\pi_{\bar M}$ and $p_{\bar M}$)
\cite{krovi2016quantum}.
We also define unitary states
\begin{align}
    \ket{\pi_M} \equiv \sum_{u \in M} \sqrt{\frac{\vec\pi_{u}}{p_M}} \ket{u},
\end{align}
and $\ket{\pi_{\bar M}}$ analogously.

To implement the interpolated quantum walk,
we use three additional registers.
The first register $\ket q$ is an ancilla qubit used to implement the oracle.
The second register $\ket c$ is an ancilla qubit used to control the interpolation.
And the third register $\ket{i}$ is the target qubit of the interpolation.
We define the oracle by
\begin{align}
    Q \ket{q c i} \ket{u0} = \begin{cases}
        \ket{(q \oplus 1) c i} \ket{u0},\text{if } u \in M\\
        \ket{qci} \ket{u0}, \text{if } u \in \bar M.
    \end{cases}
    \label{eq:oracle}
\end{align}
By letting $r \equiv 1 - 2^{-c}$,
we define the interpolation operator $\I(r)$ that maps
\begin{align}
    \ket{c}\ket{0} \to \ket c (\sqrt{1 - r} \ket 0 + \sqrt r \ket 1) .
    \label{eq:I(r)}
\end{align}
We also define the controlled operators
\begin{align}
    C(U') \equiv \ket 1 \bra 1 \otimes U' + \ket 0 \bra 0 \otimes I
\end{align} 
and $\bar C(U')$, defined analogously but activated when the control is in state $\ket 0$.
We can then implement an interpolated oracle $Q(r)$ by
\begin{align}
    Q(r) \equiv Q C(\I(r) \otimes I) Q,
    \label{eq:interpolated-oracle}
\end{align}
where the control of $C(\I(r))$ is the ancilla qubit $\ket q$.
Henceforth, we omit the ancilla registers $\ket q$ and $\ket c$.
Lastly, we define $A(r)$ by
\begin{align}
    A(r) \equiv \bar C(A) Q(r),
\end{align}
where the control of $\bar C(A)$ is the target of interpolation $\ket i$.
Note that $A(r)$ takes $\ket{0}\ket{x0}$ to $\ket{0}A\ket{x0}$ if $x \in \bar M$
and $A(r)$ takes $\ket{0}\ket{x0}$ to
\begin{align}
    \sqrt{1 - r} \ket 0 A \ket{x0} + \sqrt{r} \ket 1 \ket{x0}
\end{align}
if $x \in M$.
By using $A(r)$ we define the interpolated quantum walk operator $W(r)$ by
\begin{align}
    W(r) \equiv A(r)^\dagger (I \otimes S) A(r) \bar C(R). 
\end{align}
Let
\begin{align}
    D(r) \equiv \begin{bmatrix}
        D_{\bar M \bar M} & \sqrt{1 - r}D_{\bar M M} \\
        \sqrt{1 - r}D_{M \bar M} & (1 - r)D_{M M} + rI
    \end{bmatrix}
    \label{eq:D(r)}
\end{align}
be the discriminant of $P(r)$.
Note that $W(r)$ implements $D(r)$,
\begin{align}
    \bra{0}\bra{u0} W(r) \ket 0 \ket{v0} = D(r)_{uv}.
\end{align}

\subsection{Interpolated quantum search}
\label{sec:interpolated-search}
\noindent
Ambainis et al.~proposed a quantum algorithm (\ambainis's algorithm) that
performs quantum search with success probability of $\Omega(1)$
as long as the quantum walk implements a reversible Markov chain
\cite{ambainis2020quadratic}.
The algorithm uses amplitude amplification (AA) and
quantum fast-forwarding (QFF) as subroutines
\cite{brassard2002quantum, krovi2016quantum}.
Before describing the algorithm, we quickly review AA and QFF.

For AA, it is provided an oracle $Q$ that marks $\ket{\pi_M}$ and an
algorithm that generates the state
\begin{align}
    \ket{\psi_0} \equiv \sqrt{p_M} \ket{\pi_M} + \sqrt{1 - p_M} \ket{\pi_{\bar M}} .
\end{align}
AA takes $\ket{\psi_0}$ to $\ket{\pi_M}$ in $O(1/\sqrt{p_M})$
calls to the algorithm and to the oracle with
success probability of at least $\max(p_M, 1 - p_M)$.

QFF is used to accelerate quantum walks on reversible Markov chains.
The key idea is to use the fact that
the discriminant matrix $D$ has real singular values
which can be written as cosine functions.
Consequently,
$W^t$ implements the $t$-th Chebyshev polynomial of $D$.
Then, they use the fact that $\cos(\sqrt t \theta)$
is a pointwise approximation of $\cos^t(\theta)$
to implement $\cos^t(\theta)$ as a linear combination of
$\cos(\ell\theta)$ such that
\begin{align}
    \abs{\cos^t(\theta) -
         \sum_{\ell = 0}^{\ell_\max} w_\ell \cos(\ell\theta)
    }
    \leq \eps,
\end{align}
where $\ell_\max = \ceil{\sqrt{2t\ln(2/\eps)}}$ and $w_\ell$ are some specific weights.
This approximation is implemented using linear combination of unitaries,
which requires a new register $\ket{\ell}$ with $\log(\ell_\max)$ additional qubits
\cite{childs2012hamiltonian, berry2015simulating, van2017quantum}.
Given a state $\ket\psi$,
QFF returns the state
$\ket{D^t \psi} + \ket{\perp}$ in $O(\sqrt t)$ steps where
$\ket{D^t \psi}$ is $\eps$-close
to $D^t \ket\psi$,\footnote{Two
    states $\ket a$ and $\ket b$ are $\eps$-close if
    $\norm{\ket a - \ket b} \leq \eps$.}~ and
$\ket\perp$ is orthogonal to $\ket{D^t \psi}$.
To summarize, QFF approximately implements the discriminant of
any reversible Markov chain quadratically faster.

\begin{algorithm}[htbp]
\caption{\ambainis's algorithm.}
\label{alg:ambainis}
\begin{algorithmic}[1]
    \Require $HT$ and oracles for $P$ and $M$.
    \State \label{step:start}
        Let $t_\max = 72HT$,
        and $c_\max = \ceil{\log(36 t_\max)}$.
    \State \label{step:setup}
          Setup state
          \begin{align}
              \sum_{t = 1}^{t_\max} \sum_{c = 0}^{c_\max} \frac{1}{\sqrt{t_\max(c_\max + 1)}}
              \ket{t}\ket{c}\ket{\pi}.
          \end{align}
    \State Apply the oracle $Q$ once and measure its target (ancilla) qubit.
        If $\ket{1}$, measure the last register and output the marked vertex.
        Otherwise, we are left with state
        \begin{align}
            \sum_{t = 1}^{t_\max} \sum_{c = 0}^{c_\max}
            \frac{1}{\sqrt{t_\max(c_\max + 1)}}\ket{t}\ket{c} \ket{\pi_{\bar M}}
        \end{align}
    \State \label{step:end}
        Set $\eps = O(1/\log(t_\max))$ and use QFF controlled on
        the first two registers to map
        \begin{align}
            \ket t \ket{c} \ket{\pi_{\bar M}} \to
            \ket t \ket{c} \pr{ \ket{D(r)^t \pi_{\bar M}} + \ket\perp}.
        \end{align}
    \State Apply AA to steps~\ref{step:start} to~\ref{step:end}
        $O\pr{\sqrt{\log(t_\max)}}$ times,
        taking the success probability to $\Omega(1)$.
\end{algorithmic}
\end{algorithm}

With AA and QFF in hands,
we state \ambainis's algorithm in Alg.~\ref{alg:ambainis}.
In Alg.~\ref{alg:ambainis}, $\ket c$ is
the register used to control the interpolation
(see Eq.~\ref{eq:I(r)}),
and $\ket t$ is a \emph{new} register used to control QFF.
The additional registers $\ket q$ and $\ket i$ used to implement $D(r)$, and
the additional register $\ket{\ell}$ used to implement QFF are omitted.

In Theorem~\ref{theo:ambainis},
we restate the cost of \ambainis's algorithm and
the scenarios where it works.

\vspace*{12pt}
\begin{theorem}
    \label{theo:ambainis}
    If $P$ is a reversible Markov chain and $p_M \leq 1/9$,
    Alg.~\ref{alg:ambainis} finds a marked vertex with
    success probability $\Omega(1)$ in
    \begin{align}
        O\pr{
            \sqrt{\log(HT)}\pr{
                \mathbb{S} +
                \mathbb{W}\sqrt{HT \log\log(HT)}
            }
        }
    \end{align}
    steps where
    $HT$ is the hitting time of $P$,
    $\mathbb{S}$ is the setup cost of step~\ref{step:setup}, and
    $\mathbb{W}$ is the cost of invoking $W(r)$
    (which includes the cost of update operation and
    the cost of querying the oracle).
\end{theorem}

\vspace*{12pt}
\noindent
Note that if $p_M > 1/9$,
we can make $O(1)$ calls to the oracle before running the algorithm to
find a marked vertex with success probability of $\Omega(1)$.
A similar algorithm was proposed if $HT$ is not known in advance.

The proof of Theorem~\ref{theo:ambainis} is strongly dependent of
the interpolated Markov chain $P(r)$.
The success probability of the algorithm is lower bounded by the probability of
\begin{enumerate*}[label=(\roman*)]
    \item starting from $\vec\pi_{\bar M}$;
    \item after $t$ steps of $P(r)$ reaching a marked vertex;
    \item after additional $t'$ steps of $P(r)$ reaching an unmarked vertex.
\end{enumerate*}
If $r$ is too small we may leave the marked vertex too soon and
if $r$ is too large we may leave it too late.
This problem is tackled by taking a superposition of $\ket c$
(which leads to a superposition of $D(r)$).
This suggests that for each graph, there exists a value of $r$
for which the algorithm would work with no need of the superposition.
This was conjectured in the original paper but,
to the best of our knowledge,
it is yet to be proven.

For the remainder of this paper,
we also omit registers $\ket t$ and $\ket c$.

\subsection{Staggered quantum walk}
\label{sec:st}
\noindent
The Staggered Quantum Walk (\st) was proposed by Portugal et al.
\cite{portugal2016stqw}.
The key idea is to cover all possible transitions (edges) using cliques.
These cliques are used to define the evolution operator.
We now define the necessary concepts.

Let $G(V, E)$ be a graph.
A clique is a complete subgraph -- not necessarily maximal.
A tessellation $\Tau$ is a partition of $V$ into cliques.
For all $\tau \in \Tau$,
the subgraph $G[\tau]$ is a clique with edges $E(G[\tau])$.
The set of edges of a tessellation
$E(\Tau) \equiv \bigcup_{\tau \in \Tau} E(G[\tau])$
is a subset of $E(G)$.
Normally, $E(\Tau)$ is a proper subset of $E(G)$.
A tessellation cover is a set of tessellations $\{\Tau_j\}$ such that
$\bigcup_j E(\Tau_j) = E(G)$.
A graph is $k$-tessellable if at least $k$ tessellations are sufficient to
obtain a tessellation cover.

Throughout this paper,
we only focus on $2$-tessellable graphs and
their tessellation covers $\set{\Tau_1, \Tau_2}$.
We label
the cliques in $\Tau_1$ by $\alpha_1, \ldots,$ $\alpha_{\abs{\Tau_1}}$, and
the cliques in $\Tau_2$ by $\beta_1, \ldots, \beta_{\abs{\Tau_2}}$.
Let $\G'$ be a 2-tessellable graph.
\st s generalize \sz s because \sz s only take place on the edges of balanced bipartite graphs, whereas \st s take place on the vertices of 2-tessellable graphs. The set of 2-tessellable graphs is larger than the set of balanced bipartite graphs as follows from those two results:
\begin{enumerate*}[label=(\roman*)]
    \item~Portugal~\cite{portugal2016stqwongraphs} has shown that a graph $\G'$ is 2-tessellable if and only if its clique graph $\G=K(\G')$ is 2-colorable; and
    \item~Peterson~\cite{PETERSON2003223} has shown that the clique graph $K(\G')$ is 2-colorable if and only if $\G'$ is the line graph of a bipartite multigraph, $\G' = L(\G)$.
\end{enumerate*}
Therefore, $\G'$ is 2-tessellable if and only if it is the line graph of a bipartite multigraph.

Let $V_j$ be the $j$-th part of a bipartite multigraph $\G$.
We label the multiedges as $uve$, where
$u \in V_1$, $v \in V_2$, and
$e$ is an additional label used to distinguish
multiedges that are incident to both $u$ and $v$.
We use the same labels for elements in $E(\G)$ and $V(L(\G))$.
We emphasize that if $u \in V_j(\G)$,
then $u$ induces a clique in $\Tau_j$ by
considering the vertices in $L(\G)$ that correspond to
all multiedges in $\G$ incident to $u$.
Thus, according to the previous clique labelling,
for all $u \in V_1$,
\begin{equation}
\alpha_u = \set{uve \mid uve \in E(\G) , \forall\ v \text{ and } e},
\end{equation}
and for all $v \in V_2$,
\begin{equation}
   \beta_v = \set{uve \mid uve \in E(\G) , \forall\ u \text{ and } e}.
\end{equation}
The tessellations are given by $\Tau_1=\{\alpha_u \mid u\in V_1\}$
and $\Tau_2=\{\beta_v \mid v\in V_2\}$.

Furthermore, given a 2-tessellable graph $L(\G)$ and
a tessellation cover $\set{\Tau_1, \Tau_2}$,
we can apply an inverse process using the clique graph $K(L(\G))$ to reconstruct
the bipartite multigraph~$\G$.

To define the evolution operator,
we  define auxiliary states $\ket{\alpha_u}$ and $\ket{\beta_v}$.
For every $\alpha_u \in \Tau_1$, we define
\begin{align}
    \ket{\alpha_u} = \sum_{uve \in \alpha_u} a_{uve} \ket{uve},
\end{align}
and for every $\beta_v \in \Tau_2$, we define
\begin{align}
    \ket{\beta_v} = \sum_{uve \in \beta_v} b_{uve} \ket{uve},
\end{align}
where
$a_{uve}$ and $b_{uve}$ are complex numbers such that
$\abs{a_{uve}}, \abs{b_{uve}} > 0$, and
$\ket{\alpha_u}$ and $\ket{\beta_v}$ are unitary.
Note that $\set{\ket{\alpha_u}}$ form an orthonormal basis --
this is also true for $\set{\ket{\beta_v}}$.
We have used the same labels for \st s and \sz s in
other to highlight their relationship.
In the general case, $\ket{\alpha_u}$ and $\ket{\beta_v}$ are used to
construct \emph{Hamiltonians}, which are used to obtain the evolution operator
\cite{portugal2017stqwhamiltonian}.
However, we focus on the subcase where
the evolution operator coincides with the definition of Eq.~\ref{eq:sz-evol-op}.

\begin{figure}[!htb]
    \vspace*{13pt}
    \centering
    \begin{subfigure}[t]{0.3\textwidth}
        \centering
        \resizebox{\textwidth}{!}{\begin{tikzpicture}
    \begin{scope}[every node/.style={draw,circle}]
        
        \node (34) at (-1, -1) {34};
        \node (31) at (-2, -2) {31};
        \node (32) at (-2, 0) {32};
        \node (42) at (-1, 1) {42};
        \node (12) at (-2, 2) {12};
        
        \node (24) at (1, -1) {24};
        \node (21) at (2, -2) {21};
        \node (23) at (2, 0) {23};
        \node (43) at (1, 1) {43};
        \node (13) at (2, 2) {13};
    \end{scope}
    
    \begin{scope}[draw=red,ultra thick]
        \draw (12) -- (13);
        \draw (21) -- (23); \draw (23) -- (24); \draw (24) -- (21);
        \draw (31) -- (32); \draw (32) -- (34); \draw (34) -- (31);
        \draw (42) -- (43); 
    \end{scope}
    
    \begin{scope}[draw=blue,dashed,ultra thick]
        \draw (21) -- (31);
        \draw (12) -- (32); \draw (32) -- (42); \draw (42) -- (12);
        \draw (13) -- (23); \draw (23) -- (43); \draw (43) -- (13);
        \draw (24) -- (34); 
    \end{scope}
\end{tikzpicture}}
        \caption{$L(G[P])$}
        \label{fig:line-bip-graph}
    \end{subfigure}
    \hspace{0.1\textwidth}
    \begin{subfigure}[t]{0.3\textwidth}
        \centering
        \resizebox{\textwidth}{!}{\begin{tikzpicture}
    \begin{scope}[every node/.style={draw,circle}]
        \node [red, ultra thick] (44) at (0, 0) {\white{44}};
        \node [blue, ultra thick, dashed] (44p) at (0, 0) {\black{44}};
        
        \node [blue, ultra thick, dashed] (34) at (-1, -1) {\black{34}};
        \node (31) at (-2, -2) {31};
        \node (32) at (-2, 0) {32};
        \node [red, ultra thick] (42) at (-1, 1) {\black{42}};
        \node (12) at (-2, 2) {12};
        
        \node [blue, ultra thick, dashed] (24) at (1, -1) {\black{24}};
        \node (21) at (2, -2) {21};
        \node (23) at (2, 0) {23};
        \node [red, ultra thick] (43) at (1, 1) {\black{43}};
        \node (13) at (2, 2) {13};
    \end{scope}
    
    \begin{scope}[draw=red,ultra thick]
        \draw (12) -- (13);
        \draw (21) -- (23); \draw (23) -- (24); \draw (24) -- (21);
        \draw (31) -- (32); \draw (32) -- (34); \draw (34) -- (31);
    \end{scope}
    
    \begin{scope}[draw=blue,dashed,ultra thick]
        \draw (21) -- (31);
        \draw (12) -- (32); \draw (32) -- (42); \draw (42) -- (12);
        \draw (13) -- (23); \draw (23) -- (43); \draw (43) -- (13);
    \end{scope}
\end{tikzpicture}}
        \caption{$L(G[P'])$}
        \label{fig:line-marked-bip-graph}
    \end{subfigure}
    \vspace*{13pt}
    \caption{Line graphs and tessellations of previous bipartite graphs.
        Cliques $\alpha_u$ are induced by the red solid edges and vertices,
        and cliques $\beta_v$ are induced by blue dashed edges and vertices.}
\end{figure}
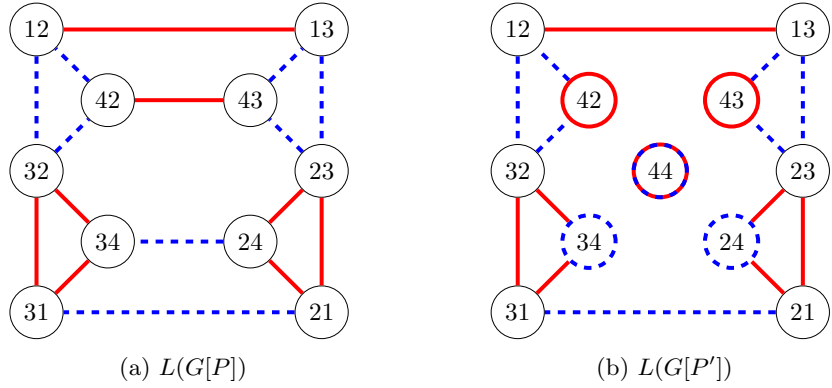

We now illustrate the relationship between \sz s and \st s.
Fig.~\ref{fig:line-bip-graph} is the line graph of
the bipartite graph depicted in Fig.~\ref{fig:bip-graph} with
tessellation cover $\set{\Tau_1, \Tau_2}$.
Cliques in $\Tau_1$ are induced by the solid red edges and vertices,
while cliques in $\Tau_2$ are induced by the dashed blue edges and vertices.
For example, $\alpha_4 = \set{42, 43}$ and $\beta_4 = \set{24, 34}$.
Fig.~\ref{fig:line-marked-bip-graph} is the line graph of
the biparte graph depicted in Fig.~\ref{fig:marked-bip-graph} with
tessellation cover $\set{\Tau'_1, \Tau'_2}$.
Since we marked vertex 4 on the Markov chain,
$\alpha'_4 = \beta'_4 = \set{44}$.
However, a tessellation is a partition of the vertex set,
hence we need to add vertices $42$ and $43$ to $\Tau'_1$, and
$34$ and $24$ to $\Tau'_2$.
So we add four cliques:
$\alpha'_{42} = \set{42}$ and $\alpha'_{43} = \set{43}$ to $\Tau'_1$, and
$\beta'_{34} = \set{34}$ and $\beta'_{24} = \set{24}$ to $\Tau'_2$.
Note that marking a vertex in the bipartite (multi)graph corresponds to
marking the corresponding clique in the line graph by removing its edges.
By doing so,
if the walker reaches a vertex of the marked clique, e.g. $42$ in $\Tau'_1$,
and the operator $2\proj_{\alpha'} - I$ is applied,
the walker stays in the same place.

\begin{figure}[!htb]
    \vspace*{13pt}
    \centering
    \begin{subfigure}[t]{0.23\textwidth}
        \centering
        \resizebox{\textwidth}{!}{\begin{tikzpicture}
    \begin{scope}[every node/.style={draw,circle}]
        \node (4x) at (0, 0) {4};
        \node (3x) at (0, 1) {3};
        \node (2x) at (0, 2) {2};
        \node (1x) at (0, 3) {1};

        \node (4y) at (2, 0) {4};
        \node (3y) at (2, 1) {3};
        \node (2y) at (2, 2) {2};
        \node (1y) at (2, 3) {1};
    \end{scope}

    \path (1x) edge[bend left=30] (2y);
    \draw (1x) -- (2y); \draw (1x) -- (3y);
    \draw(2x) -- (1y); \draw (2x) -- (3y); \draw (2x) -- (4y);
    \draw(3x) -- (1y); \draw (3x) -- (2y); \draw (3x) -- (4y);
    \draw (4x) -- (2y); \draw (4x) -- (3y);

    \draw [decorate, decoration={brace, amplitude=5pt}]
        (-0.5, 3.4) -- (0.5, 3.4)
        node[midway, yshift=1.25em]{$V_1$};
        
    \draw [decorate, decoration={brace, amplitude=5pt}]
        (1.5, 3.4) -- (2.5, 3.4)
        node[midway, yshift=1.25em]{$V_2$};
\end{tikzpicture}}
        \caption{$\G$.}
        \label{fig:bip-multigraph}
    \end{subfigure}
    \hspace{0.1\textwidth}
    \begin{subfigure}[t]{0.3\textwidth}
        \centering
        \resizebox{\textwidth}{!}{\begin{tikzpicture}
    \begin{scope}[every node/.style={draw,circle}]
        
        \node (34) at (-1, -1) {34};
        \node (31) at (-2, -2) {31};
        \node (32) at (-2, 0) {32};
        \node (42) at (-1, 1) {42};
        \node (12) at (-2, 2) {12};
        
        \node (24) at (1, -1) {24};
        \node (21) at (2, -2) {21};
        \node (23) at (2, 0) {23};
        \node (43) at (1, 1) {43};
        \node (13) at (2, 2) {13};

        \node (12p) at (0, 2) {$12'$};
    \end{scope}
    
    \begin{scope}[draw=red,ultra thick]
        \draw (12) -- (12p); \path (12p) edge (13); \path [bend right] (13) edge (12);
        \draw (21) -- (23); \draw (23) -- (24); \draw (24) -- (21);
        \draw (31) -- (32); \draw (32) -- (34); \draw (34) -- (31);
        \draw (42) -- (43); 
    \end{scope}
    
    \begin{scope}[draw=blue,dashed,ultra thick]
        \draw (21) -- (31);
        
        \draw (12) -- (32); \draw (32) -- (42); \draw (42) -- (12);
        \path [bend right] (12p) edge (32); \draw (32) -- (42); \draw (42) -- (12p);
        \draw (12) -- (12p);
        
        \draw (13) -- (23); \draw (23) -- (43); \draw (43) -- (13);
        \draw (24) -- (34); 
    \end{scope}
\end{tikzpicture}}
        \caption{$L(\G)$.}
        \label{fig:line-bip-multigraph}
    \end{subfigure}
    \vspace*{13pt}
    \caption{A bipartite multigraph $\G$ and its line graph $L(\G)$.
        We relabelled $uv0 \to uv$ and $uv1 \to uv'$.}
\end{figure}
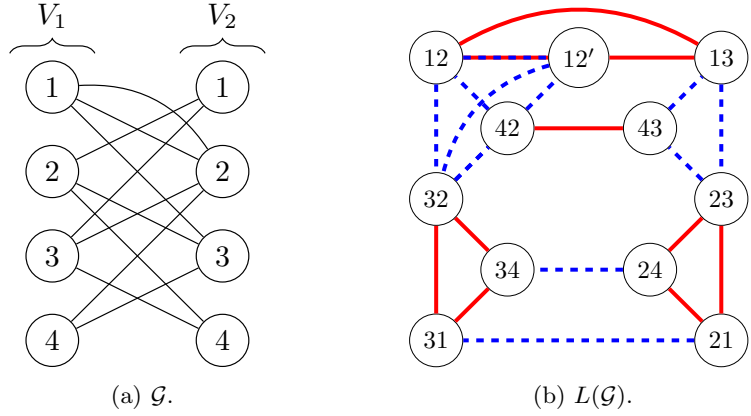

To obtain some insight of \st s' generalization of \sz s,
we add a multiedge to the graph $G[P]$ of Fig.~\ref{fig:bip-graph}.
For example, if we add multiedge $121$,
we obtain multigraph $\G$ in Fig.~\ref{fig:bip-multigraph}.
This graph cannot be obtained by the process described in Section~\ref{sec:sz}
because this process only adds a single edge per transition.
Note that multiedge 121 cannot be described in the Hilbert space defined for \sz s.
For this reason, the dynamics of $\G$ cannot be simulated exactly using \sz s.
The corresponding line graph $L(\G)$ is depicted in Fig.~\ref{fig:line-bip-multigraph}.
By adding multiedge 121, a new vertex was added to $L(\G)$.
Consequently, cliques $\alpha_1$ and $\beta_2$ were redefined.
Note that $E(L(\G)[\alpha_1]) \cap E(L(\G)[\beta_2]) \neq \emptyset$.
Whenever this occurs,
the underlying bipartite multigraph always has multiedges.

We now define the (double) discriminant matrix associated with \st s.
Define $(D_2)_{uv} \equiv \braket{\alpha_u | \beta_v}$ and $D_1 \equiv D_2^\dagger$.
With $D_1$ and $D_2$ in hands, we can define the double discriminant matrix
\begin{align}
    D \equiv \begin{bmatrix}
        0 & D_2 \\ D_1 & 0
    \end{bmatrix} .
\end{align}
$D$ is written by making a disjoint union of the indices of
$\alpha_u$ and $\beta_v$.
That is, $D$ is defined with respect to the computational basis
$\set{\ket 1, \ldots, \ket{\abs{\Tau_1} + \abs{\Tau_2}}}$,
and we associate $\ket{u}$ with $\ket{\alpha_u}$ for all $1 \leq u \leq \abs{\Tau_1}$,
and $\ket{\abs{\Tau_1} + v}$ with $\ket{\beta_v}$ for all $1 \leq v \leq \abs{\Tau_2}$.

Higuchi et al. \cite{higuchi2019eigenbasis} have
computed the eigenbasis of 2-tessellable \st s.
They defined the quantum equivalent of \cdb:
the \emph{Quantum Detailed Balance condition} (\qdb).
\qdb\ is satisfied if there exists
$\ket{\pi_j} \in \text{span}\{\ket 1, \ldots, \ket{\abs{\Tau_j}}\}$
such that for all $uve \in E(\G)$,
\begin{align}
    \braket{uve | \alpha_u} \braket{u | \pi_1} &=
    \braket{uve | \beta_v} \braket{v | \pi_2}.
    \label{eq:qdb}
\end{align}
We assume that $\norm{\ket{\pi_1}} =\norm{\ket{\pi_2}} = 1$, and
define the not normalized states
\begin{align}    \ket{\pi} &\equiv \ket{\pi_1} \oplus \ket{\pi_2},\quad\text{and}\\
    \ket{- \pi} &\equiv \ket{\pi_1} \oplus (- \ket{\pi_2}).
\end{align}
Higuchi et al. found necessary and sufficient conditions for the \qdb\ to hold.
We restate this result in Lemma~\ref{lemma:higuchi}.

\vspace*{12pt}
\begin{lemma}
    \qdb\ is satisfied if and only if
    $\ket\pi$ ($\ket{-\pi}$) is the unique
    $+1$-eigenvector ($-1$-eigenvector) of $D$.
    \label{lemma:higuchi}
\end{lemma}

\vspace*{12pt}

Higuchi et al. have also shown how to obtain \cdb\ from \qdb.
Note that $\braket{uve | \alpha_u}$ is related to the probability of going
from vertex $u$ to $v$ by multiedge $uve$ --
alternatively,
going from clique $\alpha_u$ to $\beta_v$ through vertex $uve$.
Summing the probability associated to each multiedge incident to both $u$ and $v$,
we obtain a column-stochastic matrix $P_1$ with entries
\begin{align}
    (P_1)_{vu} \equiv \sum_e \abs{\braket{uve | \alpha_u}}^2
    = \sum_e \abs{a_{uve}}^2.
\end{align}
Analogously,
\begin{align}
    (P_2)_{uv} \equiv \sum_e \abs{\braket{uve | \beta_v}}^2
    = \sum_e \abs{b_{uve}}^2.
\end{align}
Then, by taking the square modulus of Eq.~\ref{eq:qdb} and
summing over all multiedges $uve$ incident to both $u$ and $v$,
we obtain the set of equations
\begin{align}
    \sum_e \abs{\braket{uve | \alpha_u} \braket{u | \pi_1}}^2
    &=
    \sum_e \abs{\braket{uve | \beta_v} \braket{v | \pi_2}}^2
    \\
    (P_1)_{vu} (\vec\pi_1)_u &= (P_2)_{uv} (\vec\pi_2)_v.
    \label{eq:bipartite-cdb}
\end{align}
We also refer to Eq.~\ref{eq:bipartite-cdb} as \cdb\ because 
if we define a new transition matrix $P$ as a block matrix of $P_1$ and $P_2$,
\begin{align}
    P \equiv \begin{bmatrix}
        0 & P_2 \\ P_1 & 0
    \end{bmatrix},
    \label{eq:bip-P}
\end{align}
we obtain the original \cdb\ (Eq.~\ref{eq:cdb}).

Nevertheless, we also obtain a relation similar to Eq.~\ref{eq:discriminant}
using the \qdb.
Note that if we fix $u$,
\begin{align}
    a_{uve} &= b_{uve} \frac{\braket{v | \pi_2}}{\braket{u | \pi_1}}
\end{align}
is true for all $v$ and $i$.
By definition of $D_2$, 
we have
\begin{align}
    (D_2)_{uv} &= \sum_e a^*_{uve} \ b_{uve} \\
    &= \frac{\braket{v | \pi_2}^*}{\braket{u | \pi_1}^*} \sum_e \abs{b_{uve}}^2 \\
    &= \frac{1}{\braket{u | \pi_1}^*} (P_2)_{uv} \braket{v | \pi_2}^* .
\end{align}
Similar results follow for $D_1$.
Therefore, by defining $\Lambda^{-1} \equiv \diag{\ket\pi^*}$,
we obtain the desired relation\footnote{Note that
    the entries of $\Lambda$ are amplitudes,
    not probabilities.
    So there is no need for taking the square root of the entries.
}
\begin{align}
    D = \Lambda P \Lambda^{-1}.
    \label{eq:D-similar-P}
\end{align}


\section{Adapting Szegedy Quantum Walks}
\label{sec:lsz}
\noindent
Recall that \sz s take place on the edges of a
balanced bipartite graph $G[P]$ obtained through a
duplication process from Markov chain $P$.
\ambainis's algorithm finds a vertex of $G[P]$ that is
associated to a vertex of the reversible Markov chain $P$.

Let $P$ be a bipartite Markov chain as defined by Eq.~\ref{eq:bip-P} that satisfies the \cdb.
Let $G[P_1, P_2]$ be the graph representation of $P$.
Since $G[P_1, P_2]$ is not balanced in general and
$P$ is not reversible,
\ambainis's algorithm cannot be applied directly.
To address this,
we define \sz s on arbitrary bipartite graphs.
Then, we adapt \sz s to obtain the
underlying reversible Markov chain $\P \equiv P_2P_1$.
Using this adapted version of \sz s,
we can efficiently find marked vertices in $G[P_1, P_2]$.

We now rewrite the evolution operator.
We use the following auxiliary operators,
\begin{align}
    A \ket{u0} &\equiv \ket{\alpha_u}, \text{ and}\\
    B \ket{v0} &\equiv \ket{\beta_v},
\end{align}
where analogous to the alternative implementation of \sz s,
we have extended the Hilbert space of the second register with
a reference state $\ket 0$ orthogonal to vertices' labels.
$A$ implements the transitions of $P_1$,
while $B$ implements the transitions of $P_2$.
Also, we did not use the swap gate $S$ to implement $B$,
which allows us to work with the smaller Hilbert space
$\hilb^{N_1} \otimes \hilb^{N_2 + 1}$ instead of
$\hilb^{N_\max} \otimes \hilb^{N_\max + 1}$
where $N_\max = \max(N_1, N_2)$.
We define reflection operators similar to the one defined in
Eq.~\ref{eq:ref-operator},
\begin{align}
    R_j \equiv 2 \sum_{u = 1}^{N_j} \ket{u0}\bra{u0} - I.
\end{align}
Note that $R_1 = 2\proj_\alpha - I$ and $R_2 = 2\proj_\beta - I$.
We also define auxiliary operators
\begin{align}
    W_1 &\equiv B^\dagger A R_1,\ \text{and} \\
    W_2 &\equiv A^\dagger B R_2.
\end{align}
We rewrite the evolution operator as
\begin{align}
    U &= A W_2 W_1 A^\dagger \\
    &= B R_2 B^\dagger A R_1 A^\dagger \\
    &= \pr{2 \proj_\beta - I}\pr{2 \proj_\alpha - I}.
\end{align}
Note that if we take $P_2 = P_1$ and $V_1 = V_2$,
we obtain $B = SA$ and $W_1 = W_2$,
as depicted in Eq.~\ref{eq:original-sz-U}.
Also, note that $U^t = A (W_2 W_1)^t A^\dagger$.

Recall that in \ambainis's algorithm,
$W$ implements the discriminant matrix if $W$ is restricted to a subspace:
$\bra{u0}W\ket{w0} = D_{uw}$.
This is not true for the above $U$ as
\begin{align}
    \bra{u0}W_2 W_1\ket{w 0}
    &= \bra{\alpha_u} \pr{2\proj_\beta - I }\ket{\alpha_w} \\
    &= 2\sum_{v}\braket{\alpha_u | \beta_v} \braket{\beta_v | \alpha_w}
        - \delta_{uw} \\
    &= 2(D_2D_1)_{uw} - \delta_{uw}.
\end{align}
By taking the square of Eq.~\ref{eq:D-similar-P}, we obtain
\begin{align}
    \begin{bmatrix} D_2D_1 & 0 \\ 0 & D_1D_2 \end{bmatrix} 
    = \Lambda \begin{bmatrix} P_2P_1 & 0 \\ 0 & P_1P_2 \end{bmatrix} \Lambda^{-1}.
\end{align}
Hence, $\Lambda_1^{-1}\D \Lambda_1 = \P$, where
$\Lambda_j \equiv \diag{\ket{\pi_j}}^*$ and $\D \equiv D_2D_1$.
Using this identity,
we obtain that $W_2W_1$ implements the discriminant of the transition matrix
$2\P - I$, which is stochastic but may have negative entries.
In such cases,
we would be required to handle quasi-probability distributions,
introducing unnecessary complexity.

Therefore, we propose an alternative solution called
{adapted Szegedy's quantum walk} (adapted \sz).


\subsection{Adapted Szegedy's quantum walk}
\noindent
We adapt \sz s by changing the implementation of $A$.
We use a space orthogonal to subspace spanned by $\set{\ket{\alpha_u}}$ and
$\set{\ket{\beta_v}}$.
This allows us to
implement the discriminant $\D$ of
the reversible Markov chain $\P$ by
sending part of the amplitudes to this orthogonal space.

We now define the adapted version of the operators in
the previous section.
First, we define the auxiliary states
\begin{align}
    \ket{\alpha_u^\pm} \equiv \frac{\ket{\alpha_u} \pm \ket{u0}}{\sqrt 2}.
\end{align}
We define the operator $\A$ as an adapted version of $A$ by
\begin{align}
    \A \ket{u0} \equiv \ket{\alpha_u^+}.
\end{align}
The action of $\A$ on the remaining states is defined in a way
such that unitarity is preserved.
We also define operator $F$ that flips the phase if the second register is $\ket 0$ and
leaves the state unchanged otherwise,
\begin{align}
    F \ket{uv} \equiv \begin{cases}
        -\ket{u0} \text{ if } v = 0,\\
        \ket{uv} \text{ if } v \neq 0.
    \end{cases}
\end{align}
Note that $F\A \ket{u0} = \ket{\alpha_u^-}$.
We define the evolution operator in a similar fashion to the previous section,
\begin{align}
    \W_1 &\equiv B^\dagger F \A R_1, \text{ and} \\
    \W_2 &\equiv \A^\dagger B R_2.
\end{align}

We now show that $\W \equiv \W_2 \W_1$ implements the desired discriminant.
First, we show a couple of identities.
Summing
\begin{align}
    B \ket{v0} \bra{v0} B^\dagger
    &= \ket{\beta_v} \bra{\beta_v}
\end{align}
over $1 \leq v \leq N_2$ and
sandwiching by $\bra{\alpha_u^+}$ and $\ket{\alpha_{w}^-}$ yields
\begin{align}
    \bra{\alpha_u^+} \proj_\beta \ket{\alpha_{w}^-}
    = \frac 1 2 \bra{\alpha_u} \proj_\beta \ket{\alpha_{w}}
    = \frac 1 2 \D_{uw}.
\end{align}
Also,
\begin{align}
    \braket{\alpha_u^+ | \alpha_{w}^-}
    = \frac{\braket{\alpha_u | \alpha_{w}} - \braket{u0 | w0}}{2}
    = 0.
\end{align}
Using these identities, we obtain
\begin{align}
    \bra{u0} \W \ket{w0}
    =& \bra{\alpha_u^+} B R_2 B^\dagger F \ket{\alpha_{w}^+}
    \\
    =& \bra{\alpha_u^+} (2 \proj_\beta - I) \ket{\alpha_{w}^-}
    \\
    =& \D_{uw}.
\end{align}
Therefore, $\W$ implements the desired discriminant $\D$.

Recall that $\D$ is the discriminant matrix of $\P$.
In Section~\ref{sec:lst},
we show that $\P$ is a reversible Markov chain.
Thus, we can use QFF to implement its approximate dynamics quadratically faster.
In the next section,
we focus on describing the operators to implement the
\emph{interpolated adapted Szegedy's quantum walk}.


\subsection{Interpolated adapted Szegedy's quantum walk}
\noindent
Let $\P : V_1 \to V_1$ be a reversible Markov chain.
Then, the interpolated Markov chain $\P(r)$ is also reversible.
And we can use \ambainis's algorithm to find a vertex in $V_1$.
The focus of this section is to define the interpolated adapted operators,
and show that they implement the interpolated discriminant $\D(r)$.

We now define the auxiliary interpolated quantum walk operators.
We use the same oracle $Q$ (Eq.~\ref{eq:oracle}) and
its interpolated version $Q(r)$ (Eq.~\ref{eq:interpolated-oracle}) defined in
Section~\ref{sec:interpolated-sz}.
Recall that if $u \in M$, $Q(r)$ maps
\begin{align}
    \ket 0 \ket{u0} \to (\sqrt{1 - r}\ket 0 + \sqrt r \ket 1)\ket{u0},
\end{align}
and acts trivially otherwise.
Using $Q(r)$, we define
\begin{align}
    \A(r) &\equiv \bar C(\A) Q(r).
\end{align}
Note that $\A(r)$ takes $\ket 0 \ket{u0}$ to
$\ket 0 \ket{\alpha_u^+}$ if $u \in \bar M$,
and it takes $\ket 0 \ket{u0}$ to
\begin{align}
    \sqrt{1 - r}\ket 0 \ket{\alpha_u^+}  + \sqrt{r}\ket 1 \ket{u0}
\end{align}
if $u \in M$.
For conciseness, we define
\begin{align}
    \B &\equiv (I \otimes B),\\
    \F &\equiv (I \otimes F), \text{ and} \\
    \R_j &\equiv (I \otimes R_j).
\end{align}
Lastly, we define auxiliary operators
\begin{align}
    \W_1(r) &\equiv \B^\dagger \F \A(r) \R_1, \text{ and} \\
    \W_2(r) &\equiv \A(r)^\dagger \B \R_2.
\end{align}

We now check that $\W(r) \equiv \W_2(r) \W_1(r)$ implements $\D(r)$.
Let
\begin{align}
    \R_\beta &\equiv \B \R_2 \B^\dagger \\
    &= I \otimes (2\proj_\beta - I).
\end{align}
Recall from the previous section that
$\bra{\alpha_u^+} (2\proj_\beta - I) \ket{\alpha_w^-} = \D_{uw}$.
Hence,
\begin{align}
    \bra{0} \bra{\alpha_u^+} \R_\beta \ket{0} \ket{\alpha_w^-} &= \D_{uw}.
\end{align}
We also have that
\begin{align}
    \bra 0 \bra{\alpha_u^+} \R_\beta \ket 1 \ket{w0} = 0,
\end{align}
and
\begin{align}
    \bra 1 \bra{u0} \R_\beta \ket 1 \ket{w0} = -\delta_{uw}.
\end{align}
We now calculate all possible combinations of
$\bra 0 \bra{u0} \W(r) \ket 0 \ket{w0}$.
\begin{enumerate}[label=\roman*)]
    \item If $u, w \in \bar M$,
        \begin{align}
            \bra{0, u0}\W(r) \ket{0, w0}
            =& \bra{0, u0}\A(r)^\dagger \R_\beta \F \A(r) \ket{0, w0} \\
            =& \bra{0, \alpha_u^+} \R_\beta \ket{0, \alpha_w^-} \\
            =& \D_{uw}.
        \end{align}
        This yields the block $\D_{\bar M \bar M}$,
        analogous to the result of the previous section.
    \item If $u \in \bar M$ and $w \in M$,
        \begin{align}
            \bra{0, u0}\W(r) \ket{0, w0}
            =& \bra{0, \alpha_u^+} \R_\beta (\sqrt{1 - r} \ket{0, \alpha_w^-} - \sqrt r \ket{1,w0}) \\
            =& \sqrt{1 - r}\ \D_{uw}.
        \end{align}
        This yields the block $\sqrt{1 - r}\ \D_{\bar M M}$.
    \item Analogously, if $u \in M$ and $w \in \bar M$,
        \begin{align}
            \bra{0, u0}\W(r) \ket{0, w0} \implies \sqrt{1 - r}\ \D_{M \bar M} .
        \end{align}
    \item Lastly, if $u, w \in M$,
        \begin{align}
            \bra{0,u0}\W(r) \ket 0 \ket{0,w0}
            =& \bra{0, u0} \A(r)^\dagger \R_\beta \F \A(r) \ket{0, w0} \\
            =& (1 - r)\bra{0, \alpha_u^+} \R_\beta \ket{0, \alpha_w^-}
            - r\ \bra{1, u0} \R_\beta \ket{1, w0} \\
            =& (1 - r)\D_{uw} + r \delta_{uw}.
        \end{align}
        which implies $(1 - r)\D_{MM} + rI$.
\end{enumerate}
Putting everything together, we conclude that
$\W(r)$ implements $\D(r)$ (see Eq.~\ref{eq:D(r)}).

Since $\D(r)$ is the discriminant of $\P(r)$,
which is a reversible Markov chain if $0 \leq r < 1$,
we can apply \ambainis's algorithm to
perform quantum search on a predetermined part $V_1$ or $V_2$
even if $N_1 \neq N_2$.
Instead of predetermining a part with marked vertices,
we could require a controlled oracle that
marks vertices on the first (second) part if
the control qubit is set to $\ket 0$ ($\ket 1$).
In this case, we run \ambainis's algorithm twice:
the first time searching for marked vertices in the first part, and
the second time searching for marked vertices in the second part.

\begin{figure}[!htb]
    \vspace*{13pt}
    \centering
    \begin{subfigure}[t]{0.22\textwidth}
        \centering
        \resizebox{\textwidth}{!}{\begin{tikzpicture}
    \begin{scope}[every node/.style={draw,circle}]
        \node [ultra thick] (4x) at (0, 0) {\textbf{4}};
        \node (3x) at (0, 1) {3};
        \node (2x) at (0, 2) {2};
        \node (1x) at (0, 3) {1};

        \node (4y) at (2, 0) {4};
        \node (3y) at (2, 1) {3};
        \node (2y) at (2, 2) {2};
        \node (1y) at (2, 3) {1};
        \node (4p) at (2, -1) {$4'$};
    \end{scope}
    
    \begin{scope}[every path/.style={-latex}]
        \draw (2y) -- (4x); \draw (3y) -- (4x);
    \end{scope}

    \draw (1x) -- (2y);
    \draw (1y) -- (2x);
    \draw (1x) -- (3y); \draw (1y) -- (3x);
    \draw (2x) -- (4y); \draw (3x) -- (4y);
    \draw (2x) -- (3y); \draw (2y) -- (3x);
    \draw [-latex] (4y) -- (4x);

    \draw (4x) -- (4p);

    \draw [decorate, decoration={brace, amplitude=5pt}]
        (-0.5, 3.4) -- (0.5, 3.4)
        node[midway, yshift=1.25em]{$V_1$};
        
    \draw [decorate, decoration={brace, amplitude=5pt}]
        (1.5, 3.4) -- (2.5, 3.4)
        node[midway, yshift=1.25em]{$V_2'$};
\end{tikzpicture}}
        \caption{$G[P_1', P_2']$.}
        \label{fig:marked-bip-lsz}
    \end{subfigure}
    \hspace{0.1\textwidth}
    \begin{subfigure}[t]{0.3\textwidth}
        \centering
        \resizebox{\textwidth}{!}{\begin{tikzpicture}[every node/.style={draw,circle}]
    \node [ultra thick] (4) at (0, 0) {\textbf{4}};
    \node (3) at (1, 1) {3};
    \node (2) at (-1, 1) {2};
    \node (1) at (0, 2) {1};

    \draw (1) -- (2); \draw (1) -- (3); \draw [-latex] (1) -- (4);
    \draw (2) -- (3); \draw [-latex] (2) -- (4);
    \draw [-latex] (3) -- (4);

    \begin{scope}[>=latex]
        \path (1) edge [loop above] (1);
        \path (2) edge [loop left] (2);
        \path (3) edge [loop right] (3);
        \path (4) edge [loop below] (4);
    \end{scope}
\end{tikzpicture}}
        \caption{$\P(1).$}
        \label{fig:new-markov-chain}
    \end{subfigure}
    \vspace*{13pt}
    \caption{Interpretations of interpolated adapted \sz s.
        Vertex $4 \in V_1$ is marked.
        In Fig.~\ref{fig:marked-bip-lsz}, vertex $4'$ was added by
        the disjoint union $V_2 \sqcup M$.
        In Fig.~\ref{fig:new-markov-chain}, it is depicted the
        new Markov chain  $\P(1) = P_2' P_1'$.
    }
\end{figure}
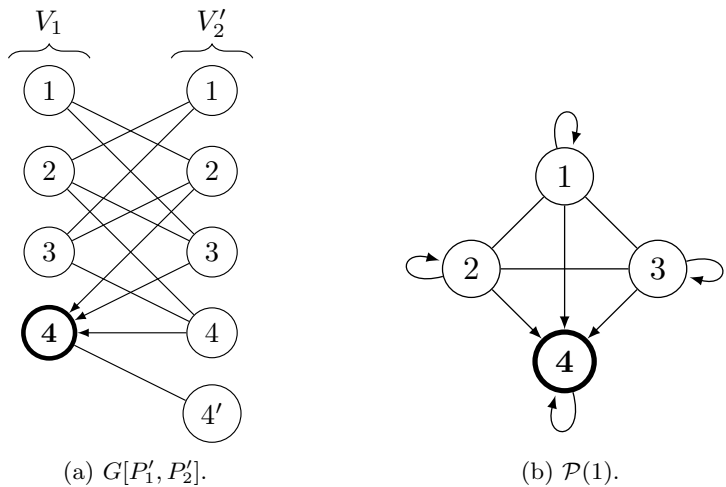

Now, we focus on the interpretations of the
actions of $\A(r)$ and $\B$.
Note that if $u \in M$,
$\A(1)$ takes $\ket 1 \ket{u0}$ to $\ket 1 \ket{u0}$.
In this case,
the walker in a marked vertex goes to a space orthogonal to the evolution of $\D(0)$.
For this reason, we interpret $\A(1)$ as implementing
a new transition matrix $P_1': V_1 \to V_2'$ where
\begin{enumerate*}[label=(\roman*)]
    \item we augment $V_2$ with $M$,
        i.e. $V_2' = V_2 \sqcup M$;
    \item we let $(P_1')_{V_2 M} = 0$,
        i.e. remove transitions $M \to V_2$;
    \item we let $(P_1')_{M M} = I$,
        i.e. adding transitions $M \to M$;
        and
    \item we leave the remaining entries unchanged.
\end{enumerate*}
Analogously, we interpret the action of $\B$ as implementing
another transition matrix $P_2' : V_2' \to V_1$ where
\begin{enumerate*}[label=(\roman*)]
    \item we let $(P_2')_{M M} = I$,
        i.e. adding transitions $M \to M$;
        and
    \item we leave the remaining entries unchanged.
\end{enumerate*}
Note that by removing transitions $M \to V_2$,
we turned the marked vertices into sinks in $\P(1)$.
Also, by adding new transitions $M \to M$,
we added self-loops to the marked vertices in $\P(1)$.
By construction, $P_1'$ and $P_2'$ are stochastic.
For example,
if we mark vertex $4 \in V_1$ in the bipartite graph $G[P]$ of Fig.~\ref{fig:bip-graph},
we obtain the bipartite digraph $G[P_1', P_2']$ of Fig.~\ref{fig:marked-bip-lsz}
and the Markov chain $\P(1)$ of Fig.~\ref{fig:new-markov-chain}.

\section{Searching with Staggered Quantum Walks}
\label{sec:lst}
\noindent

The operators from Section~\ref{sec:lsz} can be adapted to \st s by incorporating complex amplitudes and summing over all relevant elements. Specifically, for the 2-tessellable graph $L(\G)$, the state $\ket{\alpha_u}$ is obtained by summing over all vertices in the clique $\alpha_u$, while $\ket{\beta_v}$ is obtained by summing over all vertices in the clique $\beta_v$. Equivalently, for the bipartite multigraph $\G$, the state $\ket{\alpha_u}$ is constructed by summing over all multiedges incident to $u \in V_1$, and $\ket{\beta_v}$ is obtained by summing over those incident to $v \in V_2$. Applying the same modifications as in Section~\ref{sec:lsz} results in the operators of the interpolated quantum walk, yielding identical outcomes in the context of \st s.

In this section, we
\begin{enumerate*}[label=(\roman*)]
    \item explain why it was not possible to perform quantum search on \st s in the general case;
    \item show that the Markov chain $\P = P_2P_1$ (or $P_1P_2$) is reversible; and
    \item show that quantum search can be performed to find a clique (vertex) in 
        any 2-tessellable graph (bipartite multigraph).
\end{enumerate*}

Prior to this work, quantum search on 2-tessellable graphs using \st s was not possible using the techniques based on \ambainis's algorithm because \sz s take place on balanced bipartite graphs, which correspond to 2-tessellable graphs with $|\Tau_1| = |\Tau_2|$. Since $|\Tau_1| \neq |\Tau_2|$ in the general case, we would not always be able to mark $\alpha_m$ and $\beta_m$. This limitation motivated us to develop the technique presented in Section~\ref{sec:lsz}.

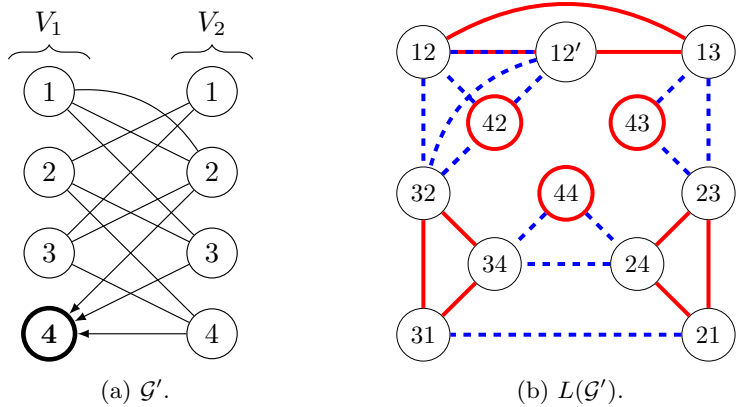
\begin{figure}[!htb]
    \vspace*{13pt}
    \centering
    \begin{subfigure}[t]{0.22\textwidth}
        \centering
        \resizebox{\textwidth}{!}{\begin{tikzpicture}
    \begin{scope}[every node/.style={draw,circle}]
        \node [ultra thick] (4x) at (0, 0) {\textbf{4}};
        \node (3x) at (0, 1) {3};
        \node (2x) at (0, 2) {2};
        \node (1x) at (0, 3) {1};

        \node (4y) at (2, 0) {4};
        \node (3y) at (2, 1) {3};
        \node (2y) at (2, 2) {2};
        \node (1y) at (2, 3) {1};
    \end{scope}
    
    \begin{scope}[every path/.style={-latex}]
        \draw (2y) -- (4x); \draw (3y) -- (4x);
    \end{scope}

    \draw (1x) -- (2y); \path [bend left] (1x) edge (2y);
    \draw (1y) -- (2x);
    \draw (1x) -- (3y); \draw (1y) -- (3x);
    \draw (2x) -- (4y); \draw (3x) -- (4y);
    \draw (2x) -- (3y); \draw (2y) -- (3x);
    \draw [-latex] (4y) -- (4x);

    \draw [decorate, decoration={brace, amplitude=5pt}]
        (-0.5, 3.4) -- (0.5, 3.4)
        node[midway, yshift=1.25em]{$V_1$};
        
    \draw [decorate, decoration={brace, amplitude=5pt}]
        (1.5, 3.4) -- (2.5, 3.4)
        node[midway, yshift=1.25em]{$V_2$};
\end{tikzpicture}}
        \caption{$\G'$.}
        \label{fig:marked-bip-lst}
    \end{subfigure}
    \hspace{0.1\textwidth}
    \begin{subfigure}[t]{0.3\textwidth}
        \centering
        \resizebox{\textwidth}{!}{\begin{tikzpicture}
    \begin{scope}[every node/.style={draw,circle}]
        \node [red, ultra thick] (44) at (0, 0) {\black{44}};
        
        \node (34) at (-1, -1) {34};
        \node (31) at (-2, -2) {31};
        \node (32) at (-2, 0) {32};
        \node [red, ultra thick] (42) at (-1, 1) {\black{42}};
        \node (12) at (-2, 2) {12};
        
        \node (24) at (1, -1) {24};
        \node (21) at (2, -2) {21};
        \node (23) at (2, 0) {23};
        \node [red, ultra thick] (43) at (1, 1) {\black{43}};
        \node (13) at (2, 2) {13};

        \node (12p) at (0, 2) {$12'$};
    \end{scope}
    
    \begin{scope}[draw=red,ultra thick]
        \draw (12) -- (12p); \path (12p) edge (13); \path [bend right] (13) edge (12);
        \draw (21) -- (23); \draw (23) -- (24); \draw (24) -- (21);
        \draw (31) -- (32); \draw (32) -- (34); \draw (34) -- (31);
    \end{scope}
    
    \begin{scope}[draw=blue,dashed,ultra thick]
        \draw (21) -- (31);
        
        \draw (12) -- (32); \draw (32) -- (42); \draw (42) -- (12);
        \path [bend right] (12p) edge (32); \draw (32) -- (42); \draw (42) -- (12p);
        \draw (12) -- (12p);
        
        \draw (13) -- (23); \draw (23) -- (43); \draw (43) -- (13);
        \draw (24) -- (34); \draw (44) -- (34); \draw (44) -- (24);
    \end{scope}
\end{tikzpicture}}
        \caption{$L(\G')$.}
        \label{fig:marked-line-bip-lst}
    \end{subfigure}
    \vspace*{13pt}
    \caption{Interpretations of quantum search in 2-tessellable graphs.
        In Fig.~\ref{fig:marked-bip-lst} vertex $4 \in V_1$ is marked.
        Equivalently, in Fig.~\ref{fig:marked-line-bip-lst},
        clique $\alpha_4$ is marked.
    }
\end{figure}

The set of 2-tessellable \st s are more general than \sz s as discussed in Section~\ref{sec:st}. Moreover, marking a vertex (creating a sink) in the bipartite multigraph (Fig.~\ref{fig:marked-bip-lst}) corresponds to removing the edges of the corresponding clique in the line graph (Fig.~\ref{fig:marked-line-bip-lst}). In the proposed setup, it is not necessary to mark vertices (cliques) in both parts (tessellations). Another reason why \ambainis's algorithm cannot be applied is the absence of a reversible Markov chain associated with \st s.

Throughout the remainder of this section,
we focus on the bipartite multigraph $\G$ as it emphasizes the relationship between \sz s and \st s.
We now show how to obtain \qdb\ from \cdb\ and vice versa
even if the associated bipartite multigraph is not connected.

\vspace*{12pt}
\begin{proposition}
    \qdb\ is satisfied if and only if \cdb\ is satisfied.
    \label{prop:qdb-iff-cdb}
\end{proposition}

\vspace*{12pt}
\begin{proof}
    Suppose \qdb\ is satisfied.
    By following the same steps of Section~\ref{sec:st},
    after Lemma~\ref{lemma:higuchi},
    we obtain the \cdb.

    Now, suppose that $(P_1)_{vu} (\vec\pi_1)_u = (P_2)_{uv} (\vec\pi_2)_v$.
    From the \cdb, we obtain a bipartite graph $G$.
    We show how to obtain a \qdb\ for any bipartite multigraph $\G$
    with underlying simple graph $G$.
    From $\vec\pi_j$ we can construct $\ket\pi$ (and $\Lambda$)
    by letting $\braket{u|\pi_j} = \exp(\im\theta_u)\sqrt{(\vec\pi_j)_u}$
    for real values of $\theta_u$.
    Construct $\ket{\alpha_u}$ by assigning complex amplitudes $a_{uve}$ to $\ket{uve}$
    such that $uve \in E(\G)$, $a_{uve} \neq 0$ and $\sum_e \abs{a_{uve}}^2 = (P_1)_{vu}$
    for all $v$.
    Note that we can use as many multiedges $uve$ as desired.
    By construction,
    $\set{\ket{\alpha_u}}$ form an orthonormal basis:
    $\braket{\alpha_{u'} | \alpha_u} = 0$ if $u \neq u'$ and
    \begin{align}
        \braket{\alpha_u|\alpha_u} &= \sum_v (P_1)_{vu} = 1.
    \end{align}
    Let the entries of $\ket{\beta_v}$ be
    $b_{uve} = a_{uve}\braket{u|\pi_1}/\braket{v|\pi_2}$ for all $u$ and $e$.
    Note that $\ket{\beta_v}$ is unitary by construction,
    \begin{align}
        \braket{\beta_v | \beta_v}
        &= \sum_{ue} \abs{a_{uve}}^2 \abs{\frac{\braket{u|\pi_1}}{\braket{v|\pi_2}}}^2 \\
        &= \sum_{u} (P_1)_{vu} \frac{(\vec\pi_1)_u}{(\vec\pi_2)_v} \\
        &= \sum_u (P_2)_{uv} = 1.
    \end{align}
    $\set{\ket{\beta_v}}$ is an orthonormal basis by construction.
\end{proof}

Another useful result is how to obtain $P_2$ if
we are given some transition matrix $P_1$,
ensuring that $P_1$ and $P_2$ satisfy the \cdb.

\vspace*{12pt}
\begin{lemma}
    For any transition matrix $P_1: V_1 \to V_2$ and
    probability distribution $\pi_1$ with no 0-entries,
    we can define a transition matrix $P_2 : V_2 \to V_1$ and
    probability distribution $\pi_2$ that satisfies the \cdb.
    \label{lemma:P1-to-P2}
\end{lemma}

\vspace*{12pt}
\begin{proof}
    Suppose that $P_1 : V_1 \to V_2$ and
    let $\vec\pi_1$ be a probability vector with no 0-entries.
    We want to satisfy the expression
    \begin{align}
        (P_1)_{vu} (\vec\pi_1)_u = (P_2)_{uv} (\vec\pi_2)_v .
    \end{align}
    Since $P_1$ is a transition matrix from $V_1$ to $V_2$,
    define $\vec\pi_2 = P_1 \vec\pi_1$.
    This implies that
    \begin{align}
        (\vec\pi_2)_v = \sum_u (P_1)_{vu} (\vec\pi_1)_u.
    \end{align}
    Then, if we take
    \begin{align}
        (P_2)_{uv} = \frac{(P_1)_{vu} (\vec\pi_1)_u}{(\vec\pi_2)_v},
    \end{align}
    the \cdb\ is satisfied.

    However, we still need to check if $P_2$ defined in this way is a
    valid transition matrix.
    Since
    \begin{align}
        \sum_u (P_2)_{uv} = \frac{\sum_u (P_1)_{vu} (\vec\pi_1)_u}{
                                  \sum_{u'} (P_1)_{vu'} (\vec\pi_1)_{u'}}
        = 1,
    \end{align}
    $P_2$ is column-stochastic, thus a valid transition matrix.
\end{proof}

Note that $P_2$ obtained from Lemma~\ref{lemma:P1-to-P2} has
the same transitions as $P_1$ but in reverse and with different probabilities.
However, the graph $G[P_1, P_2]$ is not connected in the general case
because $P_1$ may take $V_{11} \to V_{21}$ and $V_{12} \to V_{22}$
where $V_1 = V_{11} \sqcup V_{12}$ and $V_2 = V_{21} \sqcup V_{22}$.

In the following lemma,
we state the conditions to obtain a reversible $\P$.

\vspace*{12pt}
\begin{lemma}
    If \cdb\ is satisfied and $G[P_1, P_2]$ is a connected bipartite graph,
    then  $\P = P_2P_1$ is reversible.
    \label{lemma:cdb-to-reversible}
\end{lemma}

\vspace*{12pt}
\begin{proof}
    Suppose that \cdb\ is satisfied and that
    $G[P_1, P_2]$ is a connected bipartite graph.
    We have to show that $\P$ is ergodic and that
    $\P_{u'u}\vec s_u = \P_{uu'}\vec s_{u'}$ for some stationary distribution $\vec s$.

    First we show that $\P$ is ergodic.
    Note that
    \begin{align}
        \P_{u'u} = \sum_v (P_2)_{u'v}(P_1)_{vu}.
    \end{align}
    Since $G[P_1, P_2]$ is a connected graph (not digraph),
    we have that $(P_1)_{vu} \neq 0 \iff (P_2)_{uv} \neq 0$.
    Thus, $\P$ takes $u$ to itself and its 2-neighbors in $G[P_1, P_2]$.
    By continuing in this fashion, if we let $k_0$ to be
    the length of the largest path in $G[P_1, P_2]$.
    Then for all $k \geq k_0$ and $u', u \in V_1$,
    we have $\P^k_{u'u} > 0$.
    Hence $\P$ is ergodic.
    
    Now, we show that $\P$ also respects the \cdb. Note that
    \begin{align}
        \P_{u'u} (\vec\pi_1)_u &= \sum_v (P_2)_{u'v}(P_1)_{vu} (\vec\pi_1)_u
        \\
        &= \sum_v (P_2)_{uv} (P_2)_{u'v} (\vec\pi_2)_v
        \\
        &= \sum_v (P_2)_{uv} (P_1)_{vu'} (\vec\pi_1)_{u'}
        \\
        &= \P_{uu'}(\vec\pi_1)_{u'} .
    \end{align}
    Therefore, we just need to take $\vec s = \vec\pi_1$.
\end{proof}

The implication of this lemma is that we associate
adapted \sz s and adapted \st s with a reversible Markov chain
defined in $V_1$ or $V_2$ (tessellation $\Tau_1$ or $\Tau_2$).
This lemma also gives the following corollary.

\vspace*{12pt}
\begin{corollary}
    If $\P = P_2P_1$ is reversible,
    then $\vec\pi_1$ is its unique stationary distribution.
\end{corollary}

\vspace*{12pt}
\begin{proof}
    Follows directly from the fact that $P_2P_1$ is ergodic.
\end{proof}

Note that this corollary is coherent with Lemma~\ref{lemma:higuchi}.
We did not use Lemma~\ref{lemma:higuchi} directly because
if a Markov chain $P$ has stationary distribution $\vec\pi$,
it does not necessarily imply that $P$ is reversible.

With all these results in hand,
we show that we can perform quantum search to find a vertex (clique) on
any connected bipartite multigraph (2-tessellable graph).

\vspace*{12pt}
\begin{theorem}
    Let $\G$ be a connected bipartite multigraph and
    $Q$ be an oracle that marks at most $1/9$ of the vertices in $V_1$.
    We can construct $P_1$ and $P_2$ and find a marked vertex in
    $\tilde O (\sqrt{HT})$ queries to $Q$ and success probability $\Omega(1)$,
    where $HT$ is the hitting time of $\P = P_2P_1$.
    \label{theo:main}
\end{theorem}

\vspace*{12pt}
\begin{proof}
    Let $\G$ be a connected bipartite multigraph and $Q$ be the oracle.
    $\G$ has parts $V_1$ and $V_2$ and let $G$ be its underlying simple graph.
    We use the edges of $G$ to define a transition matrix $P_1 : V_1 \to V_2$ where
    $(P_1)_{vu} \neq 0$ if and only if $u$ and $v$ are adjacent, and
    a probability vector $\vec\pi_1$ with no 0-entries.
    We let $\vec\pi_1$ be the uniform probability distribution,
    consequently $p_M \leq 1/9$.
    From Lemma~\ref{lemma:P1-to-P2}, we can construct $P_2$ such that \cdb\ is satisfied.
    We use the construction of Proposition~\ref{prop:qdb-iff-cdb} to satisfy \qdb\ and
    obtain states $\set{\ket{\alpha_u}}$ and $\set{\ket{\beta_v}}$ with
    non-zero amplitudes for each multiedge of of $\G$.
    Then, from Lemma~\ref{lemma:cdb-to-reversible},
    we know that $\P$ is a reversible Markov chain.
    By applying the interpolated adapted \st,
    the operator $\W(r)$ implements $\D(r)$,
    the discriminant matrix of $\P(r)$.
    Using Theorem~\ref{theo:ambainis}, we finish the proof.
\end{proof}

A consequence of Theorem~\ref{theo:main} is that
whoever implements the oracle does not need to know
the transition matrices $P_1$ and $P_2$.
However, $|V_1|$ must be known because the oracle marks vertices.
Since bipartite graphs are bipartite multigraphs with no multiedges, we can apply Theorem~\ref{theo:main} to perform quantum search on
any bipartite graph.
It is also very straightforward how to perform quantum search on
any connected 2-tessellable graph.

\vspace*{12pt}
\begin{corollary}
    Let $L(\G)$ be a connected 2-tessellable graph with
    previously known tessellation cover $\set{\Tau_1, \Tau_2}$.
    Let $Q$ be an oracle that marks at most $N_1/9$ cliques in $\Tau_1$.
    We can construct $P_1$ and $P_2$ and find a marked clique in
    $\tilde O (\sqrt{HT})$ queries to $Q$ and success probability $\Omega(1)$,
    where $HT$ is the hitting time of $\P = P_2P_1$.
\end{corollary}

\vspace*{12pt}
\begin{proof}
    Given a 2-tessellable graph $L(\G)$ and
    a tessellation cover $\set{\Tau_1, \Tau_2}$,
    we can apply a process using the clique graph $K(L(\G))$ to
    construct the bipartite multigraph~$\G$.
    Then, the result follows from Theorem~\ref{theo:main}.
\end{proof}


\section{Final Remarks}
\label{sec:conclusion}

In this work, we propose a quantum search algorithm for bipartite graphs using an adapted version of Szegedy's quantum walks and for 2-tessellable graphs using an adapted version of staggered quantum walks. Our approach generalizes \ambainis’s quantum search algorithm, originally designed for balanced bipartite graphs, to arbitrary bipartite multigraphs. By formulating quantum search in terms of reversible Markov chains and their corresponding discriminant matrices, we demonstrate that our technique achieves a quadratic speedup over classical Markov chain-based search methods, using AGJK’s algorithm as a subroutine.

The main contribution of this work is the extension of \ambainis’s algorithm to the class of 2-tessellable graphs by employing the adapted version of staggered quantum walks. The new algorithm encompasses the class of graphs originally addressed by \ambainis’s algorithm, namely balanced bipartite graphs.

For future work, it would be interesting to extend our approach to $k$-tessellable graphs with $k>2$. The main challenge would be developing an appropriate underlying Markov chain, as the quantum walk would take place on a graph that has no bipartite structure. It remains unclear whether quantum search would still achieve a quadratic speedup for such a broad class of graphs.

\section*{Acknowledgements}
The authors would like to thank
Alexander Rivosh,
Franklin de Lima Marquezino, and
Raqueline Azevedo Medeiros Santos for the productive discussions.
The work of G.A. Bezerra was supported by
CNPq grant number 146193/2021-0, and
CAPES grant number 88881.934368/2024-01.
The work of A. Ambainis was supported by
Latvian Quantum Initiative under European Union Recovery and Resilience Facility project no. 2.3.1.1.i.0/1/22/I/CFLA/001.
The work of R. Portugal was supported by
FAPERJ grant number CNE E-26/200.954/2022,
and CNPq grant numbers 304645/2023-0 and 409552/2022-4.

\bibliographystyle{plain}
\bibliography{bib}

\end{document}